\documentstyle[aps,prl,epsf,floats]{revtex} 

\draft

\begin{document}
\twocolumn[\hsize\textwidth\columnwidth\hsize\csname
@twocolumnfalse\endcsname 

\title{Stringent neutron-star limits on large extra dimensions}

\author{Steen Hannestad}
\address{NORDITA, Blegdamsvej 17, 2100 Copenhagen, Denmark}

\author{Georg~G.~Raffelt} 
\address{Max-Planck-Institut f\"ur Physik 
(Werner-Heisenberg-Institut), 
F\"ohringer Ring 6, 80805 M\"unchen, Germany}

\date{12 December 2001}

\maketitle
                    
\begin{abstract}
 Supernovae (SNe) are copious sources for Kaluza-Klein gravitons which
 are generic for theories with large extra dimensions.  These massive
 particles are produced with average velocities $\simeq\,0.5\,c$ so
 that many of them are gravitationally retained by the SN core.  Every
 neutron star thus has a halo of KK gravitons which decay into
 $\nu\bar\nu$, $e^+e^-$ and $\gamma\gamma$ on time scales $\simeq
 10^9$ years.  The EGRET $\gamma$-flux limits ($E_\gamma\simeq
 100~{\rm MeV}$) for nearby neutron stars constrain the
 compactification scale for $n=2$ extra dimensions to $M\agt500~{\rm
 TeV}$, and $M\agt30~{\rm TeV}$ for $n=3$.  The requirement that
 neutron stars are not excessively heated by KK decays implies
 $M\agt1700~{\rm TeV}$ for $n=2$, and $M\agt60~{\rm TeV}$ for $n=3$.
\end{abstract}

\vskip2.0pc]    


{\it Introduction.}---Theories with large extra dimensions are a new
alternative to solve the hierarchy problem of particle
physics~\cite{Arkani-Hamed:1998rs,Antoniadis:1998ig,%
Arkani-Hamed:1999nn,Han:1999sg,Giudice:1999ck}.  Apart from
cosmology~\cite{Hall:1999mk,Hannestad:2001nq}, the most restrictive
limits on the size of the extra dimensions derive from astrophysical
arguments. The existence of Kaluza-Klein gravitons, particles with an
essentially continuous spectrum of masses, is a generic feature of the
new theory. Even though the KK states couple with the strength of
ordinary gravitons, the large number of modes implies that stars are
copious sources for these particles~\cite{Cassisi:2000hy}.  Until
recently, the most restrictive limit derived from the requirement that
the SN~1987A neutrino signal was not unduly shortened by the new
energy loss, i.e.~that the fraction of energy carried away by KK
gravitons obeys $f_{\rm KK}\alt
0.5$~\cite{Cullen:1999hc,Barger:1999jf,Hanhart:2001er,Hanhart:2001fx}.
For $n=2$ or 3 extra dimensions, this constraint translates into
limits on the compactification scale $M$ summarized in
Table~\ref{tab:limits}. (One extra dimension is already excluded by
other arguments.)

We recently obtained much more restrictive limits when taking KK
graviton decays into account~\cite{Hannestad:2001jv}. Most of the KK
states are produced with masses near the kinematical threshold.  For a
SN core with $T\simeq 30~{\rm MeV}$ this implies $m\simeq 100~{\rm
MeV}$ so that the only decay channels are $\nu\bar\nu$, $e^+e^-$ and
$\gamma\gamma$ with $\tau_{2\gamma} = \frac{1}{2}\tau_{e^+e^-} =
\tau_{\nu \bar\nu} \simeq 6 \times 10^{9}~{\rm yr} \, (m/100~{\rm
MeV})^{-3}$ \cite{Han:1999sg}. The cosmic $\gamma$-ray background
measured by the EGRET satellite then yields the limits shown in
Table~\ref{tab:limits}, using all SNe in the universe as sources.  Of
course, this argument depends on the assumption that there are no fast
invisible decay channels into other KK excitations (this could be the
case for non-toroidal compactification~\cite{Kaloper:2000jb}),
and also that the graviton emission is not suppressed as in some
models~\cite{Dvali:2001gm}.

\begin{table}[b]
\caption{\label{tab:limits}Constraints from SNe and neutron stars.}
\medskip
\begin{tabular}{lcccc}
Observation \& Object&
\multicolumn{2}{c}{$f_{\rm KK}^{\rm max}$}&
\multicolumn{2}{c}{$M^{\rm min}$ [TeV]}\\
&$n=2$&$n=3$&$n=2$&$n=3$\\
\hline
\noalign{\smallskip}
\multicolumn{4}{l}{Neutrino signal}\\
\quad SN~1987A~\protect\cite{Cullen:1999hc,Barger:1999jf,%
Hanhart:2001er,Hanhart:2001fx}
&0.5&0.5&31&2.75\\
\multicolumn{4}{l}{EGRET $\gamma$-ray limits}\\
\quad Cosmic SNe~\protect\cite{Hannestad:2001jv}
&$0.5\times10^{-2}$&$0.5\times10^{-2}$&84&7\\
\quad Cas A&$1.6\times10^{-2}$&$0.6\times10^{-2}$&73&7\\
\quad PSR J0953+0755&$4.4\times10^{-5}$&$1.8\times10^{-5}$&300&19\\
\quad RX J185635--3754&$1.0\times10^{-5}$&$0.4\times10^{-5}$&454&27\\
\multicolumn{4}{l}{Neutron star excess heat}\\
\quad PSR J0953+0755& $0.5 \times 10^{-7}$ & $0.5 \times 10^{-7}$
& 1680 & 60 \\
\multicolumn{4}{l}{GLAST $\gamma$-ray sensitivity}\\
\quad RX J185635--3754&$1\times10^{-7}$&$0.5 \times 10^{-7}$&1300&60\\
\end{tabular}
\end{table}

We presently show that even more restrictive constraints follow from
the EGRET $\gamma$-ray limits for young nearby SN remnants and nearby
neutron stars.  Taking Cas~A ($3.4~{\rm kpc}$) as a first example, the
limits are comparable to the cosmic case.  The key ingredient is that
even 320~years after the explosion the decay photons come from the
direction of Cas~A within the EGRET angular resolution.  For SN~1987A
the limits are less restrictive because of its relatively large
distance of 50~kpc.

The kinetic energy of the thermally produced KK gravitons is small so
that a large fraction of those produced in the hot inner SN core
remain gravitationally trapped. Therefore, every neutron star is
surrounded by a halo of KK gravitons which is dark except for the
decays into $\simeq 100~{\rm MeV}$ neutrinos, $e^+e^-$ pairs and
$\gamma$-rays. Since neutron stars are observed as close as 60~pc, one
gains a huge flux factor relative to Cas~A.

These arguments imply that future $\gamma$-ray telescopes such as
GLAST could be in a position to find a signature for KK graviton
decays from nearby neutron stars.

However, this possibility is marginal in view of yet more restrictive
limits. The KK decays in and around neutron stars provide a heat
source which prevents it cooling below a level where its thermal
emission is comparable to the KK heating. The low measured luminosity
of some pulsars then provides the most restrictive limits on large
extra dimensions.

{\it Supernova remnant Cas~A.}---This object probably corresponds to
Flamsteed's SN of 1680. The CHANDRA \hbox{x-ray} satellite has
unambiguously observed a non-pulsing thermal x-ray source in Cas~A
\cite{Chandra}, a compact remnant, so that Cas~A was indeed formed by
a core-collapse event. At a distance of 3.4~kpc it is the closest of
the historical core-collapse SNe except for the Crab ($2~{\rm kpc}$)
which, however, is a strong EGRET source and as such not useful to
place $\gamma$-ray limits on KK decays.

The EGRET experiment has made a full sky survey for $\gamma$-ray
sources in the 30~MeV--10~GeV range~\cite{EGRET}.  No source was
detected close to the site of Cas~A, implying a point-source flux
limit of~\cite{EGRET2}
\begin{equation}\label{eq:pointlimit}
\phi_{E>100 {\rm MeV}}\alt 10^{-7}~{\rm cm}^{-2}~{\rm s}^{-1}.
\end{equation}
Photons in this energy band would come from graviton decays with
$m\agt 200~{\rm MeV}$. Assuming a SN core temperature of 30~MeV, KK
gravitons in this mass range would be produced with an average
velocity of $\simeq\,0.5\,c$, and those which escape the gravitational
potential would end up being much slower. Therefore, in 320 years the
KK graviton cloud would have expanded no further than about 150~lyr,
or an angular extent of $1.5^\circ$. Since the EGRET beam width at
these energies is about~$5^\circ$, the KK cloud is indeed equivalent
to a point source.

The expected $\gamma$-ray flux from this cloud is written as
\begin{eqnarray}\label{eq:casa}
\frac{d\phi}{dz}&=&2 f_{\rm KK} \Gamma(z) 
\frac{E_{\rm TOT}}{\langle E_{\rm KK} \rangle}
\frac{dN}{dz} \frac{1}{4 \pi d^2}\nonumber\\
&=&1.4 \times 10^{-2}~{\rm cm}^{-2}~{\rm s}^{-1}\,
f_{\rm KK}\,d_{\rm kpc}^{-2}\,T_{30}^2 \beta^{-1}
z^3\frac{dN}{dz},
\end{eqnarray}
where $z=E_\gamma/T$ with $T$ the average temperature of the
protoneutron star and $\Gamma(z)$ the decay rate of gravitons with the
mass $m=2 E_\gamma$.  $E_{\rm TOT}$ is the total energy emitted by the
SN, taken to be $3 \times 10^{53}~{\rm erg}$, and $f_{\rm KK}$ the
fraction emitted in the form of KK gravitons.  $\langle E_{\rm KK}
\rangle$ is the mean KK energy, $\frac{dN}{dz}$ the normalized photon
distribution, and $d$ the distance to the source.  Further,
$T_{30}=T/30~{\rm MeV}$ and $\beta=\langle E_{\rm KK}\rangle/T$, where
$\beta=4.25$ for $n=2$ and $\beta=5.42$ for $n=3$.  We neglect the
kinetic energy of the KK gravitons so that $E_\gamma = m/2$, and we
also neglect the Lorentz factor in the decay lifetime. These
approximations have no significant impact on our results.  The
relevant emission process is nucleon bremsstrahlung $N+N\to N+N+{\rm
KK}$~\cite{Hanhart:2001er}.

Assuming an average temperature $T=30~{\rm MeV}$ for the SN core, the
$\gamma$ luminosity of the KK cloud of Cas~A is about $f_{\rm
KK}\,1.0\times 10^4 \,L_\odot$ for $n=2$ and $f_{\rm KK}\,2.0\times
10^4\,L_\odot$ for $n=3$. The expected $\gamma$ flux at Earth is
\begin{equation}
\phi_{E>100 {\rm MeV}}=f_{\rm KK}\times
\cases{6.0 \times 10^{-6}~{\rm cm}^{-2}~{\rm s}^{-1}&for $n=2$,\cr
1.6 \times 10^{-5}~{\rm cm}^{-2}~{\rm s}^{-1}&for $n=3$.\cr}
\end{equation}
Comparing this with the EGRET limit of Eq.~(\ref{eq:pointlimit})
constrains $f_{\rm KK}$ and the compactification scale at a comparable
level to the cosmic SNe (Table~\ref{tab:limits}).

Our bounds on the compactification scale are quite robust because they
vary with the KK emission limit as
\begin{equation}\label{eq:scaling}
M^{\rm min}\propto \left(f_{\rm KK}^{\rm max}\right)^{-1/(n+2)}\,.
\end{equation}
Therefore, $M^{\rm min}$ depends only very weakly on the exact value
of the EGRET flux limit or on the assumed SN properties.

{\it Neutron Stars.}---The gravitational potential at the surface of a
neutron star with mass $M$ and radius $R$ is $\Phi_R=-G_{\rm N}
M/R=-0.154\,(M/M_\odot)\,(10\,{\rm km}/R)$. The escape velocity from
this location is
\begin{equation}
v_{\rm esc}=\sqrt{-2\Phi_R}=
0.55\,\left(\frac{M/M_\odot}{R/10\,{\rm km}}\right)^{1/2}.
\label{eq:vesc}
\end{equation}
This is similar to the average speed of the thermally produced KK
gravitons so that a SN core retains a large fraction of these
particles within a halo of size a few $R$. This halo will continue to
shine in $\gamma$-rays from KK decay even a long time after the
original SN explosion.

\begin{figure}[b]
\begin{center}
\vspace*{-0.5cm}
\epsfysize=7truecm\epsfbox{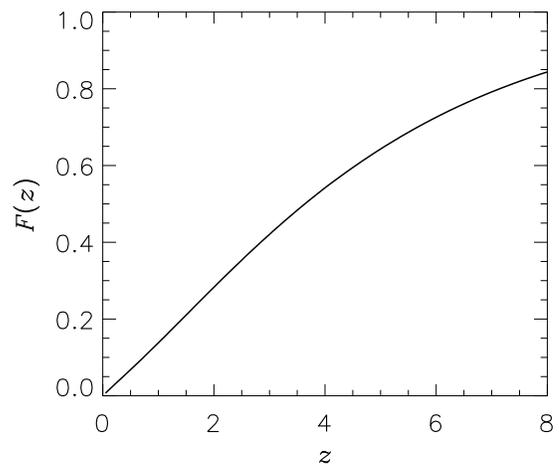}
\end{center}
\vspace*{-0.5cm}
\caption{The trapping fraction of KK-gravitons, $F(z)$, shown as a
function of $z$.}
\label{fig:F(z)}
\end{figure}

In order to calculate the expected flux we use Eq.~(\ref{eq:casa}),
taking into account two additional factors.  One is $F(z)$, the
fraction of KK-gravitons trapped close to the neutron star, which
depends on the KK-mass and thus on $z$. To calculate $F(z)$ we used $R
= 12$ km and $M = 1.4 M_\odot$, corresponding to an escape velocity of
0.59$c$. In Fig.~\ref{fig:F(z)} we show $F(z)$ as a function of $z$.
Note that it depends on $z$, but not on $n$. The differential
energy-loss rate as a function of $z=E_\gamma/T$ was shown in Fig.~1
of~\cite{Hannestad:2001jv}. Depending on $n$ it peaks for $z=2$--4,
implying an average retention fraction of about 1/2.

A second effect is that for old pulsars many of the gravitons have
decayed already, reducing the present-day source strength. Therefore,
the second factor is $\exp(-t_{\rm NS}/\tau(z))$ with $\tau(z)$ the
total lifetime of the relevant KK mode and $t_{\rm NS}$ the age of the
neutron star.

Based on this reasoning we show in Table~\ref{tab:pulsars} the limits
on $f_{\rm KK}$ obtained from several nearby neutron stars. The
closest and youngest case, RX~J185635--3754, is a non-pulsing, thermal
x-ray source.  The parallax of the optical counterpart has been
measured, yielding a distance of 60~pc. The closest EGRET source to
this neutron star is 3EG~J1847-3219 at an angular distance of about
$6.2^\circ$, far enough away to allow for useful $\gamma$-ray
limits. We have not used PSR J0437--4715~\cite{0437} ($d=0.14~{\rm
kpc}$) because it is very old so that much of the original KK
population would have decayed already.

\begin{table}
\caption{\label{tab:pulsars}Constraints on $f_{KK}$ from nearby
neutron stars.}
\begin{tabular}{llcccc}
Object&Ref.&  $d$ & $t_{\rm NS}$ & 
\multicolumn{2}{c}{$f_{\rm KK}^{\rm max}\times 10^5$}\\
&&[kpc]&$[10^6$ yr]&$n=2$&$n=3$\\ \tableline
RX J185635--3754&\cite{xray}&0.06&1.2&0.96&0.36\\
PSR J0108--1431&\cite{tauris} & 0.1 & 160&11.&4.8 \\ 
PSR J0953+0755 &\cite{0953}    & 0.12 & 17 & 4.4&1.8 \\ 
PSR J1932+1059 &\cite{1932} & 0.17 & 3.1 & 7.6 &3.0\\
\end{tabular}
\end{table}

In all cases we have assumed the EGRET point-source limit of
Eq.~(\ref{eq:pointlimit}).  In practice, the exact limit depends on
many factors, notably the source location relative to the galactic
plane and thus on the diffuse background in the neighborhood of the
relevant object. However, we really aim at a limit on the
compactification scale which depends only weakly on $f_{\rm KK}$---see
Eq.~(\ref{eq:scaling}).  Moreover, the limits from all of these
neutron stars are comparable so that one need not rely on any
particular case for an approximate overall constraint. For the two
most restrictive cases we give our nominal limits on the
compactification scale in Table~\ref{tab:limits}. These are by far the
most restrictive limits on large extra dimensions.

\begin{figure}[ht]
\begin{center}
\vspace*{-0.5cm}
\epsfysize=6.5truecm\epsfbox{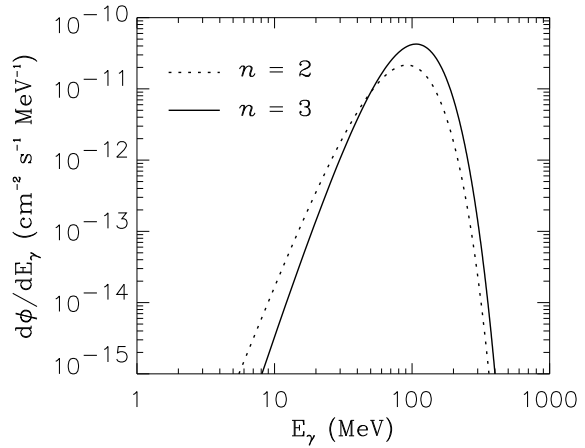}
\end{center}
\vspace*{-0.5cm}
\caption{The $\gamma$-flux from RX~J185635-3754 for $f_{\rm
KK}=10^{-7}$ and an assumed SN core temperature of 30~MeV.}
\label{fig:rx}
\end{figure}

The GLAST $\gamma$-ray satellite~\cite{glast}, to be launched in 2006,
will have a point-source flux sensitivity of $\simeq 1.5 \times
10^{-9}~{\rm cm}^{-2}~{\rm s}^{-1}$.  For the neutron star
RX~J185635--3754 this corresponds to a detection limit of $f_{\rm KK}
\simeq 10^{-7}$ for $n=2$ and ${}\simeq 0.5\times10^{-7}$ for
$n=3$. Using $f_{\rm KK}=10^{-7}$ we show in Fig.~\ref{fig:rx} the
expected $\gamma$-ray flux if the average SN core temperature was
30~MeV.

{\it Neutron star heating by KK-decays.}---Several older, isolated
neutron stars seem to have surface temperatures much higher than
expected in standard cooling models~\cite{psc96,ll99}.  The pulsars
PSR J0953+0755 and PSR J1932+1059 have both been observed with the HST
and radiation has been detected which is interpreted as almost thermal
emission from the neutron star surface. In particular for
PSR~J0953+0755 the surface temperature is tightly constrained to be $T
= (7 \pm 1) \times 10^ 4$~K.  Standard cooling models predict a much
lower temperature for a neutron star of this age.  Several models have
been proposed to explain this excess heating which appears to be of
order $10^{-5} L_\odot$~\cite{ll99}.  One promising explanation
involves friction between the crust of the neutron star and the
superfluid interior as an internal heat source.

\begin{figure}[b]
\begin{center}
\epsfysize=6.5truecm\epsfbox{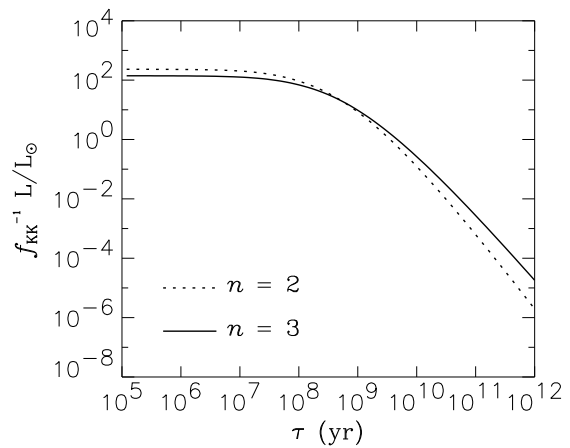}
\end{center}
\caption{The heating luminosity of a neutron star due to KK-decays as
a function of age.}
\label{fig:heating}
\end{figure}

However, the excess heat can also be generated by the cloud of
KK-gravitons surrounding the neutron star. A flux of $\gamma$-rays,
electrons, positrons, and neutrinos continuously hit the neutron star
and heat it.  (Note that a neutron star is not transparent for
neutrinos in the 100~MeV range. Note also that the charged particles
may be channelled to the polar caps by magnetic fields.) In
Fig.~\ref{fig:heating} we show the estimated heating rate from
KK-decays as a function of the neutron star age. The heating rates
plotted in the figure only include heating by gamma-rays, not
electrons, positrons, and neutrinos, so that the true heating rate
could be higher by a factor of a few.
PSR~J0953+0755 is
quite young and the energy deposited on the neutron star is therefore
200--$300\,L_\odot f_{\rm KK}$.  For the heating not to exceed the
observed luminosity one would need to require $f_{\rm KK} \alt 5
\times 10^{-8}$, by far the most restrictive limit. We stress that the
exact number is uncertain within a factor of a few, both
because the calculation of neutron star heating by KK-decays involves
some uncertainties, and because the temperature measurement of the
neutron star could be more uncertain than the quoted error bars.

Interestingly, this result implies that if neutron-star heating is
indeed due to KK-decays, then the relevant level of KK emission from a
SN core is where the GLAST satellite may just about observe KK decays
directly from RX~J185635--3754.

We have not studied in detail the cases for $n\geq4$ because we have
relied on the SN emission calculations
of~\cite{Hanhart:2001er,Hanhart:2001fx} for $n=2$ and 3. Simple
scaling arguments suggest that for $n=4$ the neutron-star heating
limit would be $M_4^{\rm min}\simeq 4~{\rm TeV}$ and $M_5^{\rm
min}\simeq 0.8~{\rm TeV}$.

{\it Summary.}---A large fraction of the KK gravitons produced by a SN
core are gravitationally retained in a cloud surrounding the neutron
star. The decay time of these particles is of order $10^9$ years, the
dominant modes have masses of order $100~\rm MeV$.  Therefore, neutron
stars should shine brightly in 100~MeV $\gamma$ rays, contrary to
observations with the EGRET satellite, allowing us to derive the most
restrictive limits on large extra dimensions. In addition, the excess
heating of neutron stars caused by the KK cloud prevents neutron stars
from cooling below a certain temperature. The observed low luminosity
of neutron stars provides even more restrictive limits on the
compactification scale.

For $n=2$ our present limits $M>500$--1600~TeV put the
compactification scale far above the weak unification scale. If large
extra dimensions solve the hierarchy problem, $M$ should not exceed
10--100~TeV, a requirement which already excludes the $n=1$ case. Our
new limits suggest that the $n=2$ case is also no longer plausible,
and even $n=3$ now seems less appealing.  Of course, our limits for
$n>1$ only apply for a situation where all extra dimensions have the
same compactification radius. In the general case the bounds could be
much less restrictive.


{\it Acknowledgments.}---In Munich, this work was partly supported by
the Deut\-sche For\-schungs\-ge\-mein\-schaft under grant No.\ SFB 375
and the ESF network Neutrino Astrophysics. We thank Thomas Janka for
calling our attention to the neutron star RX~J185635--3754 and Thomas
Tauris for discussions on nearby neutron stars. We thank Lars
Bergstr\"om and Gia Dvali for helpful comments on the manuscript.


\end{document}